\documentclass{appolb}
\usepackage{graphicx}
\usepackage{enumitem}
\begin{document}
\title{Event shape analysis in ultrarelativistic
nuclear collisions.%
\thanks{Presented at XI Workshop on Particle Correlations and Femtoscopy, Warsaw 2015}%
}
\author{Renata Kope\v{c}n\'a$^{1}$, Boris Tom\'a\v{s}ik$^{1,2}$
\address{$^1$ FNSPE, Czech Technical University in Prague, 
B\v{r}ehov\'a 7, 115~19 Praha 1, Czech Republic}
\address{$^2$ Univerzita Mateja Bela, Tajovsk\'eho 40, 
974~01 Bansk\'a Bystrica, Slovakia}
}
\maketitle
\begin{abstract}
We present a novel method for sorting events. So far, single variables like flow vector magnitude were used for sorting events. Our approach takes into account the whole azimuthal angle distribution rather than a single variable. This method allows us to determine the good measure of the event shape, providing a multiplicity-independent insight. We discuss the advantages and disadvantages of this approach, the possible usage in femtoscopy, and other more exclusive experimental studies.
\end{abstract}

  
\section{Introduction}
Initial conditions in heavy ion collisions fluctuate from event to event: there are different impact parameters and different 
initial energy-density distributions. Hot matter created in those collisions expands very fast in both longitudinal and transverse 
directions, initial inhomogeneities are translated into all orders of anisotropy 
of this expansion. The analysis of event shapes can help us 
identify events with similar initial conditions undergoing similar evolution. We present a novel study of event shapes using the 
algorithm proposed in~\cite{kopecna_jackson}. This algorithm studies the \emph{shape} of the distribution rather than a single 
variable. It compares, sorts and selects events according to \emph{similarity} with each other.

%%%%%%%%%%%%%%%%%%%%%%%%%%%%%%%%%%%%%%%%%%%%

\section{The method}
The method is thoroughly described in~\cite{kopecna_kopecna}. Here it will be briefly described using 
a simple example. % We applied the algorithm on azimuthal angle distributions. 
We generated 5000 events from a toy model. It generates azimuthal angles of pions from the distribution
\begin{equation}
\label{e:pd}
P_5(\phi) = \frac{1}{2\pi} \left (  1 + 
\sum_{n=1}^5 2v_n \cos(n(\phi - \psi_n)) 
\right )\,   .
\end{equation}
The parameters $v_n$ are quadraticaly multiplicity dependent, details can be found in ~\cite{kopecna_kopecna}, multiplicity $M \in (300,3000)$. This choice is motivated by the LHC data~\cite{kopecna_atlas,kopecna_alice}.%Those parameters for each $n$ can be found in Table~\ref{kopecna_t:params}.

%%%%%%%%%%%%%%%%%%%%%%%%%%%%%%%%%%%%%%%%%%%%%%%%%%%%%%%
% \begin{table}
% \caption{Parameters used for generating multiplicity dependent $v_n$.
% \label{kopecna_t:params}}
% \begin{center}
% \begin{tabular}{cccc}
% \hline
% $n$ & $a/10^{-8}$ & $b/10^{-5}$ & $c$ \\
% \hline\hline
% 1 & 0 &  0.01667 & --0.000680 \\
% 2 & --7.098 & 20.06 & 0.07874 \\
% 3 & --2.083 &  6.658 & 0.0424  \\
% 4 & --96.38 &  2.621 & 0.04897 \\
% 5 & --71.76 &  2.236 & 0.01673 \\
% \hline
% \end{tabular}
% \end{center}
% \end{table}
%%%%%%%%%%%%%%%%%%%%%%%%%%%%%%%%%%%%%%%%%%%%%%%%%%%%%%%

For each event, we made an azimuthal angle histogram with 20 bins. Every event is then described by its record $\{n_i\}$. 
Since we are studying \emph{angle} distribution, the choice of rotating single events is free. We will address this issue in the next section. The algorithm operates as follows \cite{kopecna_kopecna}:
\begin{enumerate}[noitemsep]
\item (Somehow rotate the events) %\vspace{2pt}
\item Order events according to a chosen variable %\vspace{2pt}
\item Divide the sorted events into quantiles (deciles) \vspace{2pt}
\item For every event calculate the probability that event with record $\{ n_i\}$ belongs to event bin $\mu$: $P(\mu|\{n_i\})$ \vspace{2pt}
\item For every event calculate mean bin number $\bar{\mu}$ (values 1 - 10):\\ $\bar{\mu} = \sum \mu P(\mu|\{n_i\})$ \vspace{2pt}
\item Sort events according to $\bar{\mu}$ \vspace{2pt}
\item If the new sorting changed the assignment of any events into event bins, return to (3). Otherwise the algorithm converged.
\end{enumerate}
Events with a similar shape are organized by the algorithm so that they end up close together. There is no specific observable according to which the sorting proceeds. Moreover, the final arrangement of events is independent of the initial sorting.

%%%%%%%%%%%%%%%%%%%%%%%%%%%%%%%%%%%%%

\section{Results}

\begin{figure}[thb]
\centerline{\includegraphics[width=0.44\textwidth]{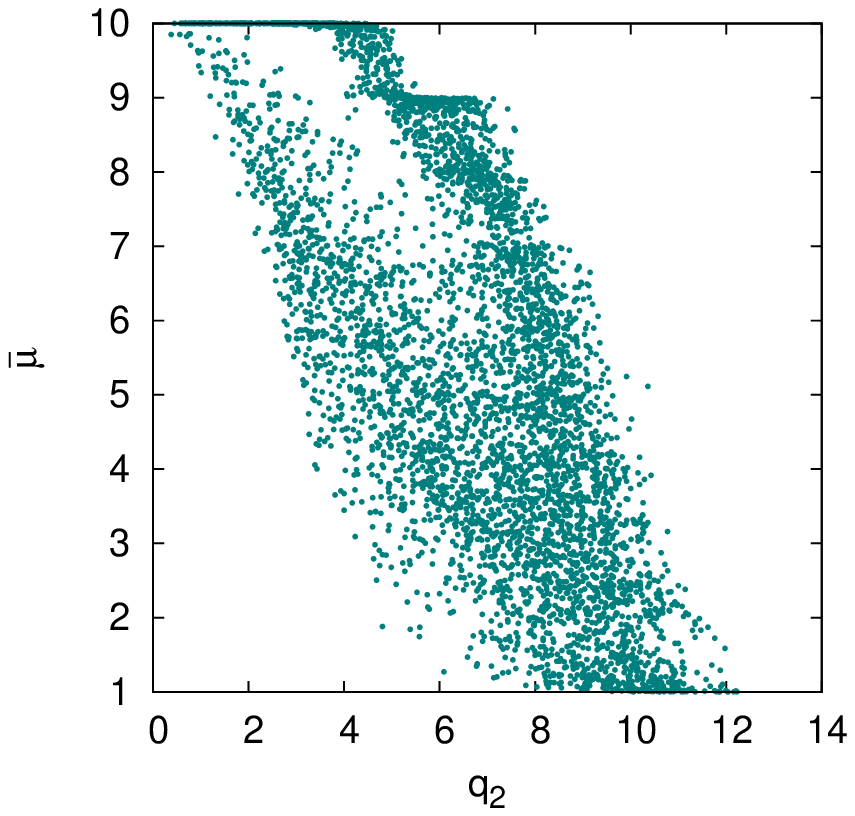}
\includegraphics[width=0.44\textwidth]{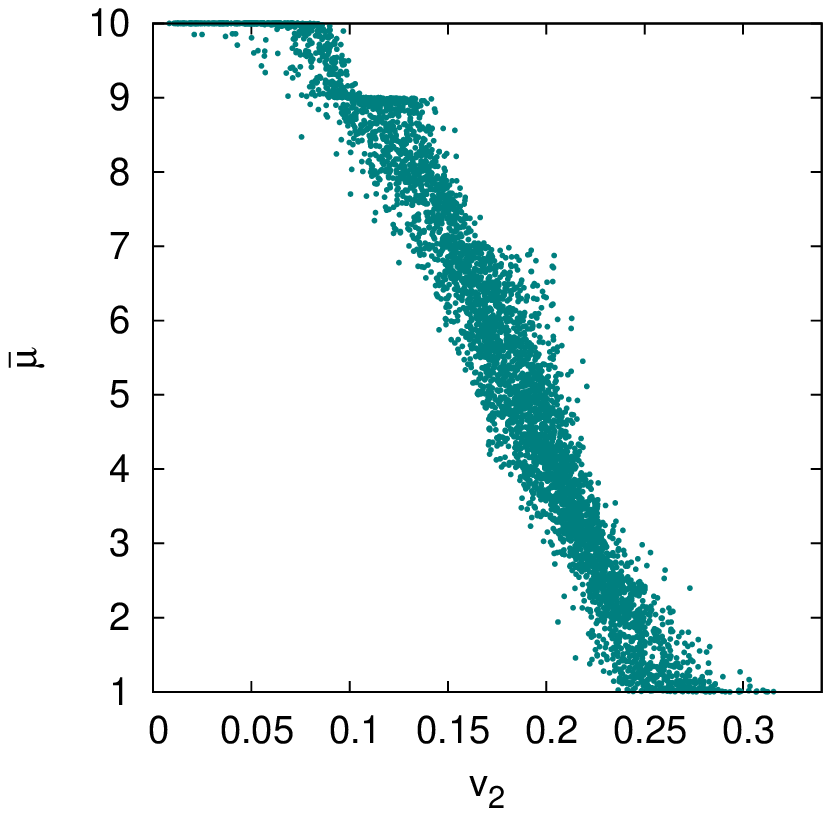}}
\caption{%
Left:  correlation of $\bar \mu$ with $q_2$ determined for each event.
Right: correlation of $\bar \mu$ with $v_2$ determined for each event via
the event plane method.
\label{kopecna_f:muq2muv2}}
\end{figure}

%\subsection{Elliptic flow}
First, we tested the algorithm using events which include only $v_1$ and $v_2$. 
One of the methods used in event shape studies is \emph{Event shape engineering}~\cite{kopecna_ESE}. 
This method sorts the events according to a chosen observable, usually $q_2 = | \sum^n_{j=1} e^{2i\phi_j}|/M$. 
We were interested in verifying whether $q_2$ is truly a good measure for sorting events. As can be seen in 
Fig.~\ref{kopecna_f:muq2muv2}, the correlation of  $\bar \mu$ with $v_2$ is clearly better than correlation of $\bar \mu$ with $q_2$. This means $v_2$ is better observable for sorting events than $q_2$ in this simple case.
 As mentioned before, the rotation of each event can be arbitrary. 
 Since in this simple case $v_2$ is clearly dominant, we decided to rotate events in a way that $\psi_2 = 0$.

%--------------------------------------------------------
%--------------------------------------------------------

%\subsection{Anisotropic flow}

\begin{figure*}[thb]
\centerline{\includegraphics[width=0.75\textwidth]{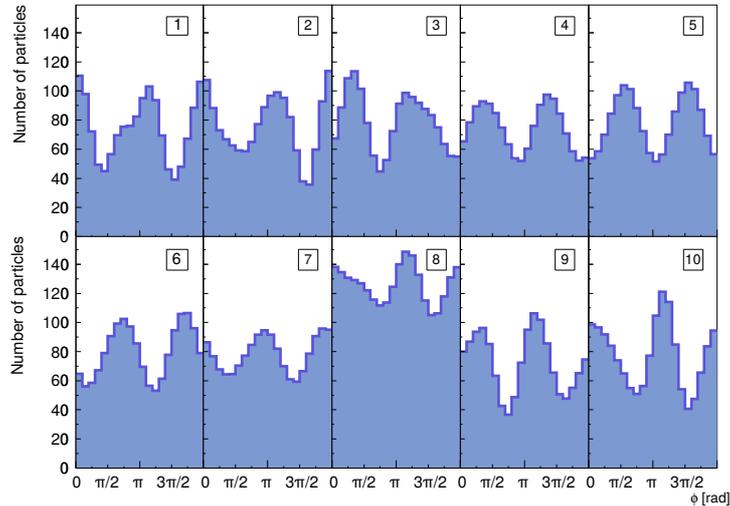}}
\caption{Average histograms of the 
azimuthal angles for event bins 1--10, with event bins indicated in the panels. 
\label{kopecna_f:histos1}}
\end{figure*}

In order to test more realistic setting, we then generated events with all five orders of Eq.~(\ref{e:pd}).
The initial event rotation is not as simple as in the previous case. Interplay of harmonics comes into play. We rotated the events according to the bisector of $\psi_3$ and $\psi_2$. Moreover, we have to take care of the parity symmetry. Hence, the events are oriented so that $\psi_2$ is less than $\pi/2$ away from $\psi_3$ counterclockwise. The final event sorting is shown in 
Fig.~\ref{kopecna_f:histos1}. It turns out that 
$v_2$ is as bad for sorting events as $q_2$ and the sorting is not even dominated by $v_3$. Higher harmonics do not play any role at all. This suggests that event shape is determined by an interplay of several observables.

%--------------------------------------------------------
%--------------------------------------------------------

% -------------------------------------------------------
% \begin{figure*}[htb]
% \centerline{\includegraphics[width=0.42\textwidth]{./figures/fig3a.eps}
% \includegraphics[width=0.42\textwidth]{./figures/fig3b.eps}}
% \centerline{\includegraphics[width=0.42\textwidth]{./figures/fig3c.eps}
% \includegraphics[width=0.4\textwidth]{./figures/fig3d.eps}}
% \caption{Correlation of various observables with final sorting variable $\bar\mu$. 
% Simulated are events with anisotropies up to 5th order and initial rotation is 
% according to $\psi_{2-3}$. Correlation with a) $v_2$, b) $v_3$, c) $q_2$, d) 
% event multiplicity.
% \label{kopecna_f:corrvn}
% }
% \end{figure*}
% -------------------------------------------------------

% \begin{figure*}
% \centerline{\includegraphics[width=0.75\textwidth]{./figures/fig4.eps}}
% \caption{Average histograms of the 
% azimuthal angles for event bins 1--10, with event bins indicated in the panels. 
% Events with anisotropies up to 5th order.
% \label{kopecna_f:allnvhist}}
% \end{figure*}

%\begin{figure}[htb]
%\centerline{%
%\includegraphics[width=12.5cm]{Fig1}}
%\caption{Plot of ...}
%\label{Fig:F2H}
%\end{figure}

%%%%%%%%%%%%%%%%%%%%%%%%%%%%%%%%%%%%%%

\section{Conclusions and outlook}

The proposed sorting algorithm provides a novel method to identify events which have evolved similarly. 
Our results confirm the importance of elliptic and triangular flows for the event shape analysis.

Our approach can be useful in studies including mixed events technique. 
One could do, e.g., a femtoscopic study of an exclusive group of events. This means that 
we could get as close as possible to singe-event femtoscopic studies. \emph{Event Shape Sorting} 
could provide a selection of events with similar momentum distributions which would make a suitable sample for 
event mixing. In case of statistics in a single event being too  small, one could take a sample of events with similar 
momentum distributions  and  reasonably  expect  that  they  also  have the  same  sizes  and  undergo  the  same  dynamics.  
Then, one could analyse the correlation function integrated over the  whole  selected  event  sample.  The  feasibility  of  
such studies will be investigated in the future.

%This method can be also used for studying U+U collisions. The asymmetric shape of U nucleus results in large fluctuations of anisotropies of the transverse flow. This algorithm could distinguish non-central tip-tip collisions from side-side or body-body collisions.

On the technical side, the required computational time is rather high, but since we have not optimised our algorithm yet, we expect the required CPU time to decrease significantly. We will also scrutinise the initial rotation of events. 

Furthermore, we are currently studying a set of events obtained by AMPT. This will bring an insight into more realistic events. 
In spite of these difficulties, we believe that our method  is worth applying in real data 
and that it will bring more detailed understanding of heavy-ion collisions dynamics.

\section*{Acknowledgement}
Supported in parts by SGS15/093/OHK4/1T/14 (Czech Republic), 
APVV-0050-11, and VEGA 1/0469/15 (Slovakia)

%%%%%%%%%%%%%%%%%%%%%%%%%%%%%%%%%%%%%%%

\end{document}